\documentstyle[12pt]{article}
\def\be{\begin{equation}}
\def\ee{\end{equation}}
\def\bea{\begin{eqnarray}}
\def\eea{\end{eqnarray}}
\newcommand{\CR}{\nonumber \\}

\newcommand{\pa}{\partial}

\def\<{\langle}
\def\>{\rangle}

\begin{document}

\baselineskip=0.7cm

\renewcommand{\thefootnote}{\fnsymbol{footnote}}
\begin{titlepage}
\begin{flushright}
UTHEP-346 \\
TU-511 \\
hep-th/9609104 \\
September, 1996
\end{flushright}

\bigskip

\begin{center}
{\Large \bf  
One-instanton calculations in $N=2$ $SU(N_c)$ Supersymmetric QCD
}

\bigskip
\bigskip

Katsushi Ito\footnote{E-mail: ito@het.ph.tsukuba.ac.jp}

\medskip

{\it Institute of Physics, University of Tsukuba\\ Ibaraki 305, Japan}

\medskip

and

\medskip 

Naoki Sasakura\footnote{Present address: 
Niels Bohr Institute, Blegdamsvej 17, DK-2100, Copenhagen {\O}, Denmark}

\medskip

{\it Department of Physics, Tohoku University \\ Sendai 980-77, Japan}

\end{center}

\bigskip

\bigskip

\begin{abstract}
We study the low-energy effective theory in $N=2$ $SU(N_{c})$ 
supersymmetric QCD with $N_f\le2N_c$ fundamental hypermultiplets
in the Coulomb branch by microscopic and exact approaches.
We calculate the one-instanton correction to the modulus 
$u\equiv \< \frac12 {\rm Tr}A^2\>$ 
{}from microscopic instanton calculation.
We also study the one-instanton corrections from the exact solutions
for $N_{c}=3$ with massless hypermultiplets.
They agree with each other except for $N_f=2N_c-2$ and $2N_c$ cases.
These differences come from possible ambiguities
 in the constructions of the exact solutions or
the definitions of the operators in the microscopic theories.  

\end{abstract}
\end{titlepage}

\renewcommand{\thefootnote}{\arabic{footnote}}
\setcounter{footnote}{0}

The prepotential of the low energy effective theories of the 
$N=2$ supersymmetric gauge theories in the Coulomb phase 
have been obtained exactly by using the holomorphy and the duality 
arguments \cite{sewi}.
The prepotential receives a non-perturbative instanton correction 
in the semi-classical region \cite{SEIZERO}.
On the other hand, the microscopic instanton 
calculations in supersymmetric gauge theories \cite{ADS,NSVZrho,AMAKON}
give reliable results when the 
semi-classical approximation is valid.
Thus the comparison between these approaches provides 
a non-trivial  check on the method of the microscopic
instanton calculus as well as the assumptions in the 
derivation of the exact solutions. 
  
Such comparisons have been studied in the case 
of $N=2$ supersymmetric Yang-Mills theories  and $SU(2)$ 
supersymmetric QCDs (SQCDs) \cite{FINPOU}-\cite{SQCD}.
The results have been  consistent with the exact solutions so far,
while some discrepancies have been reported in the $SU(2)$ SQCDs
with $N_f=3,4$ \cite{SQCD}. 
These discrepancies do not seem to contradict with the 
assumptions in the derivations of the exact solutions in the sense that
they could come from the possible ambiguities in the exact solutions or 
the definitions of the quantum observables in the microscopic theories.

In this letter we study the one-instanton correction to the
prepotential in the $N=2$ $SU(N_c)$ SQCD with $N_f\le 2N_c$ fundamental 
hypermultiplets both from the microscopic and the exact viewpoints. 
In the case of $N=2$ $SU(2)$ SQCD, 
the contributions from the odd numbers of instantons
vanish due to the anomalous ${\bf Z}_{2}$ symmetry 
since the fundamental representation of $SU(2)$ is psudoreal \cite{sewi}.
But, for $N_{c}\geq 3$, one can expect that all the instanton corrections
appear in general.

We introduce an $N=1$ chiral multiplet  $\phi=(A,\psi)$ in 
the adjoint representation and an $N=1$ vector multiplet
$W_\alpha=(v_\mu,\lambda)$, which form an $N=2$ vector multiplet.
The $N=1$ chiral multiplets 
$Q_i=(q_i,\psi_{m\, i})$ and 
$\tilde{Q}_i=(\tilde{q}_i,\tilde{\psi}_{m\, i})$ 
$(i=1,\cdots N_f)$ form the $N_f$ 
$N=2$ matter hypermultiplets in the fundamental representation. 
The microscopic $N=2$ Lagrangian for $N=2$ SQCD 
is given by 
\bea
{\cal L}&=&2 \int d^4 \theta  {\rm Tr} 
\left(\phi^\dagger e^{-2gV} \phi e^{2gV}\right)
+ \frac{1}{2g^2}\left( \int d^2\theta W^\alpha W_\alpha
+ {\rm h.c.} \right) \CR
&&+\int d^4 \theta \sum_{k=1}^{N_f}
\left( Q^\dagger_k e^{-2gV} Q_k + \tilde{Q}_k e^{2gV} 
\tilde{Q}^\dagger_k \right)
+\left( i \sqrt{2} g \int d^2\theta \sum_{k=1}^{N_f}
\tilde{Q}_k \phi Q_k + h.c. \right) \CR
&&+ \left( \int d^2 \theta \sum_{k=1}^{N_f} m_k \tilde{Q}_k Q_k
+h.c. \right) ,
\label{orilag}
\eea
where $g$ is the gauge coupling constant 
and the trace is taken in the fundamental representation.
Here the color indices are suppressed.
We will examine the euclidean lagrangian of (\ref{orilag}) 
in terms of the component fields in the Wess-Zumino gauge.

The Coulomb branch of this theory is parameterized by the expectation
values of the adjoint scalar vacuum expectation values
${A_{0\, i}}^j=a_i\delta_i^j$ and ${A_{0\, i}^\dagger}^j=
\bar{a}_i\delta_i^j$.
For generic values of $a_i$ and $\bar{a}_i$ the non-abelian gauge
symmetry completely breaks down to the $U(1)^{N_c-1}$, and the system
is in the Coulomb phase.

Since the holomorphy argument \cite{AMAKON,SHIVAI}
for the gauge coupling $g$ shows that 
the calculation in the region $g \ll 1$ is enough to 
obtain reliable results, we will perform  microscopic 
instanton calculation in the lowest order of $g$.
In this approximation, the equation of motion of the 
gauge field $D_\mu G^{\mu\nu}=0$ has the instanton solutions \cite{BPST}.
Their bosonic zero modes of the one-instanton solutions
are the instanton location $x_0$ in the euclidean space, the instanton
size $\rho$ and the location in the color space.
The integration over the location in the color space is defined by the 
integration over the minimal embedding of the subgroup $SU(2)$,
where the one-instanton configuration resides, into the gauge group
$SU(N_c)$ \cite{BER}.
The generators of the minimally embedded $SU(2)$ subgroup can be
characterized by $\Omega^\dagger J^a \Omega$, where $\Omega \in SU(N_c)$, and
$J^a$ are the generators of the $SU(2)$ subgroup obtained by 
the upper-left-hand corner embedding
of the two-dimensional representation of $SU(2)$ into the 
$N_c$-dimensional representation of $SU(N_c)$ \cite{BER}. 
Hence the integration in the color space is performed by sweeping $\Omega$ in 
$V(N_c)\equiv SU(N_c)/SU(2)\times U(1) \times SU(N_c-2)$,
where the $U(1) \times SU(N_c-2)$ is the stability group of the 
embedding and the additional $SU(2)$ \cite{SHIVAINON} 
comes from the fact that
we are interested only in the observables symmetric under the space
rotation.
By the global gauge transformation, the group
integration can be performed by rotating the scalar vacuum expectation
values $\< A\> = \Omega A_0 \Omega^\dagger, \< A^\dagger\> = 
\Omega A_0^\dagger \Omega^\dagger$, while the instanton configuration
is fixed at the upper-left-hand corner \cite{ADS}. 
 
The equations of motion of the adjoint fermions are given by 
$\tau_\mu^-D_\mu\psi=\tau_\mu^-D_\mu\lambda=0$ in the lowest order of
$g$.
{}From the index theorem,
each of these equations has $2N_c$ zero modes.
Similarly, each of the lowest order equations of motions of the matter 
fermions  
$\tau_\mu^-D_\mu\psi_{m\, i}=\tau_\mu^-D_\mu \tilde{\psi}_{m\, i}=0$
has one zero mode.
We use $\xi$, $\zeta$,  $\eta$ and $\tilde{\eta}$ to label the zero-modes
of $\lambda$, $\psi$, $\psi_{m}$ and  $\tilde{\psi}_{m}$, respectively.
Then, under an appropriate choice of the normalization of these fermionic 
zero modes\footnote{The normalizations of the zero-modes are taken 
to be unity
under the norm $\int d^4x 2 {\rm Tr}(\phi^\dagger \phi)$ and 
$\int d^4x \phi^\dagger \phi$ for adjoint and fundamental fields, 
respectively.}, the integration measure of the $N=2$ one-instanton 
zero modes is given by \cite{BER,AMAKON,COR}
\bea
&&2^{10}\pi^{2N_c+2}\Lambda_d^{b_1}g^{-4N_c} \int d^4x_0 \int_0^\infty 
d\rho \rho^{4N_c-5} \int_{V(N_c)} d\Omega
\int d^{2N_c}\xi  d^{2N_c}\zeta d^{N_f}\eta d^{N_f}\tilde{\eta},\CR
&&\Lambda_d^{b_1}\equiv\mu^{b_1}\exp\left( -\frac{8\pi^2}{g^2}\right),  
\label{instmeasure}
\eea
where $\mu$ is the Pauli-Villars regulator and $b_1=2N_c-N_f$ is
the one-loop coefficient of the beta function.
 
Out of the $4N_c$ zero-modes of the adjoint fermions, four
are the supersymmetric zero-modes $\xi_{SS}^\alpha$,  $\zeta_{SS}^\alpha$
obtained by the 
supersymmetry transformations of the one-instanton configuration of the 
gauge field.
Another four are the superconformal zero-modes 
$\xi_{SC}^\alpha$,  $\zeta_{SC}^\alpha$ 
obtained by the superconformal transformations. 
The fermionic zero-modes other than the supersymmetric ones,
say $\xi'\equiv (\xi_{SC},\xi_a,\bar{\xi}_a)$ and 
$\zeta'\equiv (\zeta_{SC},\zeta_a,\bar{\zeta}_a)$ $(a=3,\cdots, N_c)$,
cease to be zero-modes by taking into account the mass terms and 
the Yukawa terms \cite{ADS,NSVZrho}.
One of the lowest order contributions may be obtained by
substituting the solution of the scalar field equation of motion 
in the lowest order $D_\mu^2 A^\dagger(x)=0$ with the 
asymptotic value $\< A^\dagger \>$,  
and the fermionic zero-modes into the Yukawa coupling term 
$g\int d^4x{\rm Tr}( \psi[A^\dagger,\lambda])$.
Thus  we obtain a bilinear term $\xi' gM(\<A^\dagger\>)\zeta' $ 
\cite{itsa} with
\be
gM(\<A^\dagger\>)= ig\left( \begin{array}{ccc}
\sqrt2 \varepsilon \< A^\dagger\>_{tl}^{(1)} & 
(\< A^\dagger\>^{(3)})^t & 
\varepsilon \< A^\dagger\> ^{(2)} \\
\< A^\dagger\> ^{(3)} & 0 & 
-{{\rm Tr} \< A^\dagger\> ^{(1)} \over \sqrt{2}} I_{N_{c}-2}
+ \sqrt{2}\< A^\dagger\> ^{(4)} \\
(\varepsilon \< A^\dagger\> ^{(2)})^t & 
-{{\rm Tr}\< A^\dagger\> ^{(1)}\over \sqrt{2}} I_{N_{c}-2}
+\sqrt{2}(\< A^\dagger\> ^{(4)})^t 
&0\\ 
\end{array}
\right),
\label{massmatrix}
\ee
where $\varepsilon$ and  $I_{N_{c}-2}$ are a two by two antisymmetric
tensor with $\varepsilon^{12}=1$  and an $N_c-2$ by  $N_c-2$ identity
matrix, respectively. Here we have divided 
the row and column of the scalar field into the following blocks;
\be
A=\left( \begin{array}{cc} 
A^{(1)} & A^{(2)} \\
A^{(3)} & A^{(4)} \\
\end{array}
\right),
\ee
where $A^{(1)}$, $A^{(2)}$, $A^{(3)}$ and $A^{(4)}$ are 
$2\times2$, $2\times (N_{c}-2)$, $(N_{c}-2)\times 2$ and 
$(N_{c}-2)\times (N_{c}-2)$ matrices, respectively, and 
$\< A^\dagger\>_{tl}^{(1)}$ is the traceless part of 
$\< A^\dagger\>^{(1)}$.
Another contribution of order $g$ is the term
\be
-\sum_{k=1}^{N_f} \left(\frac{ g i}{\sqrt2}{\rm Tr}(\< A\>^{(1)})
+m_k\right)\bar{\eta}_k \eta_k,
\label{massterms}
\ee
which comes from the Yukawa term 
$g \tilde{\psi}_m A \psi_m$. Here we have added also the contributions 
{}from the mass terms of the matter 
fermions\footnote{We treat the mass terms perturbatively. See
\cite{AMAKON} for another treatment.}. 

The mass terms among  the zero-modes (\ref{massmatrix}),
(\ref{massterms}) has a weight $\sqrt{g}$ per one fermionic zero-mode. 
Therefore there exist other contributions with the same order in the
case of SQCD.
In fact, one needs to introduce the terms with four fermionic zero-modes 
of order $g^2$.
One of such contributions comes from the Yukawa terms 
$g \tilde{\psi}_m\lambda\tilde{q}^\dagger$ and 
$g \tilde{q}\psi\psi_m$ mediated by the propagator
of $\tilde{q}$. Using the scalar propagators in the instanton backgrounds
\cite{PRO}, we obtain  
\be
\Delta_{\tilde{q}} S
=-\frac{g^2}{12\pi^2\rho^2} \sum_{k=1}^{N_f}\tilde{\eta}_k\eta_k
\sum_{a=3}^{N_c}\xi_a\bar{\zeta}_a - \frac{g^2}{32\pi^2\rho^2}
\sum_{k=1}^{N_f} \tilde{\eta_k}\eta_k \xi_{SC}\epsilon \zeta_{SC}.
\label{fourone}
\ee
The contribution $\Delta_{q}S$ mediated by the propagator of $q$ can be 
obtained in a similar manner. 
Due to $SU(2)_R$ symmetry, the sum 
$\Delta_{q}S+\Delta_{\tilde{q}} S$ is simply given by replacing 
$\xi_a\bar{\zeta}_a$ by $\xi_a\bar{\zeta}_a+\bar{\xi}_a\zeta_a$
in the first term in (\ref{fourone}), while the second term vanishes.

The other contribution of order $g^2$ comes from the Yukawa terms 
$g{\rm Tr}([\lambda, \psi] A^\dagger)$ 
and $g \tilde{\psi}_m A \psi_m $ mediated 
by the propagator of $A$, and the result is 
\be
\Delta_{A} S 
= -\frac{g^2}{24 \pi^2\rho^2} \sum_{k=1}^{N_f}\tilde{\eta}_k\eta_k
\sum_{a=3}^{N_c}(\xi_a\bar{\zeta}_a+\bar{\xi}_a\zeta_a).
\label{dstwo}
\ee
Hence, including $8\pi^2\rho^2f$ from the contribution of the kinetic
term of $A$, we obtain the classical action of the instanton
configuration as 
\bea
S=8\pi^2\rho^2f-g\xi' M \zeta'
-\sum_{k=1}^{N_f} \left(\frac{ g i}{\sqrt2}{\rm Tr}(\< A\>^{(1)})
+m_k\right)\tilde{\eta}_k \eta_k
-\frac{g^2}{2^3\pi^2\rho^2}\sum_{k=1}^{N_f}\tilde{\eta}_k\eta_k
\sum_{a=3}^{N_c} (\xi_a\bar{\zeta}_a+\bar{\xi}_a\zeta_a),
\label{instact}
\eea
where \cite{itsa}
\be
f(\< A\>,\<A^\dagger \>)=
{\rm Tr}\left(\< A^\dagger\>_{tl}^{(1)}\< A\>_{tl}^{(1)}
+{1\over 2}\< A\> ^{(2)}\< A ^{\dagger}\> ^{(2)} +{1\over 2}
\< A^{\dagger}\>^{(3)}\< A\> ^{(3)}\right).
\ee

The supersymmetric zero-modes must be canceled by the insertion of
appropriate operators. An approach  to do this is to consider
the four fermion correlation function of the classically massless modes 
of $\psi$ and $\lambda$ \cite{SEIZERO}.
But it turns out that this approach is not so convenient 
for the cases $2N_c-2 \leq N_f \leq 2N_c$, 
because it detects only a certain combination of the fourth derivatives
of the prepotential of the 
effective theory and this is always zero for the massless 
$2N_c-2 \leq N_f \leq 2N_c$ cases from the dimensional analysis.
Thus we insert the modulus 
$u\equiv\<\frac12 {\rm Tr} A^2\>$
\cite{FUCGAB,SQCD}, which has the following
direct relation to the prepotential of the effective field theory for
the massless cases \cite{Ma};
\be
\frac{ib_1u}{2\pi}=-\Lambda {\pa {\cal F}\over\pa \Lambda}=
  \sum_{i=1}^{N_c}a_i
\frac{\partial {\cal F}}{\partial a_i}
-2{\cal F}.
\label{matone}
\ee 
Here the prepotential has the following expansion in the weak coupling
region for $N_f<2N_c$:
\be
{\cal F}(a)=\frac{\tau_0}{2}\sum_{i=1}^{N_c} a_i^2
+\frac{i}{4\pi} \left(
\sum_{i<j}(a_i-a_j)^2\log\frac{(a_i-a_j)^2}{\Lambda^2}
-{N_{f}\over 2} \sum_{i=1}^{N_{c}}a_{i}^2\log {a_{i}^2\over \Lambda^2}
\right)
-{i\over 2\pi}\sum_{n=1}^{\infty} {\cal F}_n(a) \Lambda^{b_1n},
\ee
where the first and the second terms are the classical and the one-loop
parts, respectively, and the last ones are the instanton corrections.
The logarithmic parts contribute to the classical part of $u$.
Further corrections to $u$ come purely from the instanton effects:
$u=\frac12 \sum_{i=1}^{N_c}a_i^2
+\sum_{k=1}^{\infty} u_{k}\Lambda^{b_1k}$,
where $u_k=k {\cal F}_{k}$.
On the other hand, one can obtain the modulus 
$u=\frac12 \sum_{i=1}^{N_c}a_i^2
+\sum_{k=1}^{\infty} u_k^{inst.}\Lambda_{d}^{b_1k}$ from the 
instanton calculation.
In the massive case, $u_{k}^{inst.}$ depends on $a_{i}$ and $m_{k}$.
In the following, we will determine the one-instanton contributions
$u_{1}$ and $u_{1}^{inst.}$ from the exact solutions and the 
microscopic calculation, respectively.

The contribution of the supersymmetric zero-modes to the field $A$
is obtained by solving the classical equation of the motion
$D_\mu^2A_{SS}+\sqrt2 g i [\lambda_{SS},\psi_{SS}]=0$.
The solution is given by  
$A_{SS}=\frac{ig}{4\pi} \xi^\alpha_{SS} \psi_{SS\, \alpha}$,
so we obtain \cite{FUCGAB}
\be
\int d^4x_0 u= -\frac{g^2}{2^5\pi^2} \left( \frac12 \xi_{SS}^2 \right)
\left( \frac12 \zeta_{SS}^2 \right)
\ee
for the part with the supersymmetric zero-modes.

For the massless case, after the integration over the bosonic and 
fermionic zero-modes, we obtain 
\be
\Lambda_{d,N_c,N_f}^{b_1}u_1^{inst.}(N_c,N_f)
=i^{N_f}2^{-b_1/2+1}\Lambda_{d,N_c,N_f}^{b_1}U_1^{inst.}(N_c,N_f),
\label{insres}
\ee
where we have rescaled the field $\phi\rightarrow g\phi$ to 
match with the standard convention used in the exact solutions, and
\bea
U_1^{inst.}(N_c,N_f)&\equiv&2^{-5N_c+9-N_f}\pi^{-2N_c+4} 
\int_{V(\Omega)} d\Omega \sum_{k=0}^{{\rm min}[N_f,2N_c-4]}
\,_{N_f}C_k \Gamma(2N_c-2-k) \CR
&&\times \frac{({\rm Tr}(\< A\> ^{(1)}))^{N_f-k}(-\sqrt2 i)^k  
{\rm det}_k(M(\< A^\dagger\> ))}{f(\< A\>,\< A^\dagger\>)^{2N_c-2-k}}, \CR
{\rm det}_k( M(\< A^\dagger \> ) )&\equiv&
\int d^{2N_c-2}\xi' d^{2N_c-2}\zeta' \left(
\sum_{a=3}^{N_c}(\xi_a\bar{\zeta}_a+\bar{\xi}_a\zeta_a )\right)^k
\exp\left(\xi' M(\< A^\dagger\> ) \zeta'\right).
\label{fone}
\eea

Firstly we will enumerate $U_1^{inst.}(N_c,N_f)$ by estimating the
structures of the poles \cite{itsa}.
Although the integrand depends both on $a_i$ and $\bar{a}_i$, 
the holomorphy argument tells that  $U_1^{inst.}(N_c,N_f)$ should be
a function only of the holomorphic variables $a_i$.
A pole may exist when the denominator of the integrand has some zeros
in the integration region.
This condition turns out that two of the 
$a_i$ coincides, because in this case  
$\< A\>_{tl}^{(1)} =\< A\>^{(2)} =\< A\>^{(3)}=0$ 
is realized.
Let us study the structures of the poles with the highest order.
This comes from the $k=0$ term in the sum (\ref{fone})
\footnote{The explicit integration discussed below shows that the $k=0$
term results in the expected holomorphic terms as well as some unwanted
terms such as terms with logarithmic poles and non-holomorphic terms.
These terms cancel exactly with the terms from $k>0$.}.
To estimate the structure of the pole at $a_1=a_2$, let us 
introduce an infinitesimally small parameter $\epsilon$
by $a_1-a_2=\epsilon$.
Since $f=O(\epsilon)$ at $\Omega=1$, we restrict the integration 
region to the infinitesimally small region 
$\Omega=\exp(-i\sqrt{\epsilon}\omega)\in V(N_c)$ to keep $f$ to be
$O(\epsilon)$.
Then the nonlinear integration region of $\Omega$ is linearized,
and one obtains easily the following behavior of $U_1^{inst.}(N_c,N_f)$:
\be
U_1^{inst.}(N_c,N_f)\sim 
\frac{(a_1+a_2)^{N_f}}{2^{N_f}(a_1-a_2)^2\prod_{j>2}^{N_c} (a_1-a_j)^2}.
\ee
The full expression should have the similar poles at 
$a_i=a_j\, (i\neq j)$, and so we obtain the following result up to 
possible gauge invariant regular terms:
\bea
U_1^{inst.}(N_c,N_f)&=& \frac{\sum_{i<j}^{N_c} (a_i+a_j)^{N_f}
\prod_{k<l\, ;k,l\neq i,j}^{N_c} (a_k-a_l)^2 \prod_{k\neq i,j}^{N_c}
(a_i-a_k)(a_j-a_k)}{2^{N_f}\prod_{i<j}^{N_c} (a_i-a_j)^2}.
\label{fonefull} 
\eea
The gauge invariant regular term may exist only for the cases 
$N_f=2N_c-2,\, 2N_c$ from the dimensional analysis.
For $N_f=2N_c-2$, it is a constant term, while it is a term of the 
form $const.\sum_{i=1}^{N_c} a_i^2$ for the case $N_f=2N_c$.
These terms can not be fixed by the present method of estimating the 
structures of the poles, but the explicit integrations for the cases
$N_c=3$ with $N_f\leq 2N_c$ show that the regular terms in fact 
vanish\footnote{Eq.~(\ref{fonefull}) is correct also for $N_c=2$.}.
The explicit integration is rather cumbersome even for the case
$N_c=3$.
This is simplified enormously by putting the antiholomorphic variables
$\bar{a}_i$ to special values so that the integrand takes simple 
forms \cite{itsa}, because the function 
$U_1^{inst.}(N_c,N_f)$ should be independent of the antiholomorphic variables
$\bar{a}_i$.
Putting $\bar{a}_1=\bar{a}_2=1, \bar{a}_3=-2$ and taking into account
the delta functional contribution at $\Omega=1$ \cite{itsa}, we obtain
(\ref{fonefull}) for the cases $N_f\leq 2N_c$ with $N_c=3$.

We may also consider the massive cases.
We expand with the mass terms in the instanton action (\ref{instact}).
Since these terms cancel part of the matter fermionic zero-modes in the
instanton measure, we obtain
\bea
u_1^{inst.}(N_c,N_f;m)&=&\sum_{k=0}^{N_f}
t_k(m) u_1^{inst.}(N_c,N_f-k), \CR
t_k(m)&\equiv&\sum_{i_1<\cdots<i_k}m_{i_1}\cdots m_{i_k}.
\label{insresmas}
\eea

Now we will check the consistency of the above microscopic results 
(\ref{insres}), (\ref{fonefull}) and (\ref{insresmas}) with the
physical matching condition of the dynamical scales.
The physical matching condition of the dynamical scales in the 
Pauli-Villars regularization scheme is given by \cite{FINPOU}
\be
\frac{\prod_{i} m_{Q,\, i}}{\prod_{i} m_{W,\, i}}\Lambda_d^{b_1}=
{\Lambda_d'}^{b_1'}, 
\label{phymatcon}
\ee
where $\Lambda_d$ denotes the dynamical scale of the original system,
and $\Lambda'_d$ denotes that of the induced system after integrating
out the heavy modes of the vector multiplets with masses $m_{W,\, i}$'s
and the matter multiples with masses $m_{Q,\, i}$'s.
First consider the case that some of the masses of the matters, say 
$m_{Q,\, i}\, (i=1,\cdots,k)$, are very large
compared to the others. One can show easily that, taking the 
limit $\Lambda_d\rightarrow 0$ of (\ref{insresmas}) with fixing 
$\prod_{i=1}^k m_{Q,\, i}\Lambda_d^{b_1}={\Lambda_d'}^{b_1'}$
will give a similar expression  of (\ref{insresmas}) with the 
substitution  $N_f\rightarrow N_f-k$. Another check is given by the Higgs 
breaking $(N_c,N_f)\rightarrow (N_c-1,0)$ by taking the 
limit $b\rightarrow \infty$ in $a_i=a_i'-b\, (i=1,\cdots,N_c-1)$,
$a_{N_c}=(N_c-1)b$.
The heavy masses
due to the large vacuum expectation value $b$ are given 
by
$m_W=\sqrt2 N_c b$ and $m_Q=\sqrt2 i b$.
Hence the physical matching condition of the scales is given by 
\be
\Lambda_{d,N_c-1,0}^{2N_c-2}=i^{N_f}2^{N_f/2-1}b^{N_f-2}{N_c}^{-2}
\Lambda_{d,N_c,N_f}^{2N_c-N_f}.
\label{matphy}
\ee
This is consistent with (\ref{insres}) and (\ref{fonefull}) because 
of $U_1(N_c,N_f)\sim b^{N_f-2}{N_c}^{-2}U_1(N_c-1,0)$ in the 
$b\rightarrow \infty$ limit.
  
The exact solutions are determined by the hyperelliptic curve and 
the meromorphic one-form on it \cite{sewi}.
There are some proposals with non-perturbative differences consistent
with the symmetries of the system \cite{haoz,arplsh,mine}.
Firstly, we shall use the hyperelliptic curves in \cite{haoz}.
The hyperelliptic curve and the meromorphic one-form $\lambda$ for the 
$SU(N_c)$ QCD with $N_f(<2N_c)$ flavors are given as follows:
\bea
y^2&=&F(x)^2-G(x), \CR
F(x)&=&\prod_{i=1}^{N_c}(x-e_i)+\left\{
\begin{array}{cc}
0 & \mbox{for}\ N_f<N_c, \\
2^{-2}\Lambda^{b_1} \sum_{i=0}^{N_f-N_c}x^{N_f-N_c-i}
t_i(m) & \mbox{for}\ N_f\ge N_c,\\
\end{array}
\right. \CR
G(x)&=&\Lambda^{b_1}\prod_{i=1}^{N_f}(x+m_i), \CR
\lambda&=&\frac{xdx}{2\pi i y }\left(\frac{FG'}{2G}-F'\right).
\label{curveone}
\eea
The curve for the case $N_f=2N_c$ is given by the following substitution
in the above definitions:
\bea
F(x)&=&x^{N_c}+l(q)\prod_{i=0}^{N_c-2}s_{N_c-i}x^i+
2^{-2}L(q)\sum_{i=0}^{N_c}x^{i}t_{N_c-i}(m),\CR
s_k&=&(-1)^k\sum_{i_1<\cdots<i_k}e_{i_1}\cdots e_{i_k}, \CR 
G(x)&=&L(q)\prod_{i=1}^{2N_c}(x+l(q)m_i), \CR
\lambda&=&{1\over l(q)}
\frac{xdx}{2\pi i y }\left(\frac{FG'}{2G}-F'\right).
\label{curvetwo}
\eea
Here $q\equiv \exp(2\pi i \tau)=\exp (-8\pi^2/g_{\mbox{ex}}^2)$ 
and the $L(q)$ and $l(q)$ are defined by
\bea
L(q)&=&\frac{4\theta\left[\frac12\ 0\right]^4}{\theta[0\ 0]^4},\ \ 
l(q)=\frac{\theta\left[0\ \frac12\right]^8}{\theta[0\ 0]^4}, \CR
\theta[m_1\ m_2]&=&\sum_{n\in Z^{N_c-1}} \exp\left\{
2\pi i \left[\frac12(n+m_1)^t\tau (n+m_1)+(n+m_1)^tm_2\right]\right\},
\label{theta}
\eea
where $\tau_{ij}=\tau(\delta_{ij}+1)\  (i,j=1,\cdots,N_c-1)$, 
and $0$ and $\frac12$ denote
the zero vector and a vector with one of its entries being $\frac12$
and the others are zeros, respectively. 

The vacuum expectation values of the scalar field $a_i$ can be written as
periods of the one form $\lambda$ on the curve \cite{sewi}:
\be
(\lambda_i,a)=\oint_{A_i} \lambda,
\label{aiint}
\ee
where $\lambda_i$ are the fundamental weights and $A_i$ are the 
appropriate homology cycles on the curve.
The equation (\ref{aiint}) gives $(\lambda_i,a)$'s as functions of $e_i$'s.
Inverting them, one obtains the modulus $u\equiv\frac12\sum_{i=1}^{N_c}e_i^2$
in terms of $a_i$'s.
For $N_{c}=3$, the explicit form of the contour integral (\ref{aiint})
may be obtained by solving the Picard-Fuchs equations \cite{ItYa2} 
with respect to
$u$ and $v=e_{1}e_{2}e_{3}$.
In the semi-classical region, the power series type solutions for 
$N_{f}$ flavors ($1\leq N_{f}\leq 5$) takes the form
\bea
(\lambda_1,a)&=& w(\alpha_{N_{f}}, \beta_{N_{f}};x_{1},x_{2})
   +{1\over2} w(\gamma_{N_{f}}, \delta_{N_{f}};x_{1},x_{2}), \CR
(\lambda_2,a)&=& w(\alpha_{N_{f}}, \beta_{N_{f}};x_{1},x_{2})
   -{1\over2} w(\gamma_{N_{f}}, \delta_{N_{f}};x_{1},x_{2}), 
\eea
where
\be
x_{1}={v^{2}\over u^{3}}, \quad 
x_{2}=\Lambda^{6-N_{f}} 
u^{N_{f}/2-3 (3+(-1)^{N_{f}})/4} v^{-(1-(-1)^{N_{f}})/2}, 
\ee
and $\alpha_{N_{f}}= -{1-(-1)^{N_{f}}\over 4 (6-N_{f})}$,
$\beta_{N_{f}}= \delta_{N_{f}}=-{1\over 6-N_{f}}$,
$\gamma_{N_{f}}= {9+(-1)^{N_{f}} (3-2 N_{f})\over 4 (6-N_{f})}$.
$w(\alpha,\beta; x_{1},x_{2})$ denotes  a power series
of the form $\sum_{m,n\geq 0} d_{m,n} x_{1}^{m+\alpha} x_{2}^{n+\beta}$ with
$d_{0,0}=1$. The coefficients $d_{m,n}$ are determined recursively. 
For $N_{f}=6$ the explicit evaluation of the contour integral 
would be effective
instead  of  using the Picard-Fuchs equations.
For massless cases we may derive 
the one-instanton correction to $u$ for $N_c=3$ by
the explicit enumeration of the integral, which is expanded 
in powers of $\epsilon=\Lambda^{b_1/2}$. 
The calculation goes as follows. 
Let $e_i'$ and $e_i''$ denote the two branch points of the curves 
(\ref{curveone}), (\ref{curvetwo}) 
approaching to $e_i$ in the limit $\epsilon\rightarrow 0$.
When the homology cycle in the right hand side of (\ref{aiint})
is taken around the $e_i'$ and $e_i''$, one obtains directly $a_i$. 
The $e_i'$ and $e_i''$ can be expanded in the integral powers of
$\epsilon$, and we take the terms up to order $\epsilon^4$. 
We expand the  one-form $\lambda$ up to order $\epsilon^3$,
after the change of variable $x=e_i+\epsilon z$. 
Performing the contour integral explicitly, we obtain $a_i$ in terms of 
$e_j$'s.
Inverting the results, we obtain $u=\frac12 \sum_{i=1}^{N_c} e_i^2$
in terms of $a_i$'s.
Taking the term with $\epsilon^2$, we obtain the one-instanton
correction as 
\bea
\Lambda_{N_c,N_f}^{b_1}u_1&
=&2^{-1}\Lambda_{N_c,N_f}^{b_1} U_1(N_c,N_f), \CR 
U_1(N_c,N_f)&=&
\frac{\sum_{i=1}^{N_c}{a_i}^{N_f} 
\Delta^{N_c-1}(a_{1},\ldots, \widehat{a_{i}},\ldots, a_{N_c} )}
{2\Delta^{N_c}(a_{1},\ldots, a_{N_c})}
+A_{N_c}\delta_{N_f,2N_c-2} 
+B_{N_c}\delta_{N_f,2N_c}\sum_{i=1}^{N_c}{a_i}^2,\CR
\Delta^{m}(a_{1},\ldots, a_{m})&\equiv&\prod_{k<l}^m (a_{k}-a_{l})^2 
\label{fonar}
\eea
for the cases $N_f\leq 2N_c$ with $N_c=3$, where $A_3=0$.
For the case $N_f=2N_c$ with $N_c=3$, the scale parameter
$\Lambda_{N_c,N_f}^{b_1}$ should be replaced by the $L(q)\sim 64q$,
and we obtain $B_3=-\frac78$ by using $l(q)\sim 1-40q$.

We have also calculated the modulus $u$ in other curves. For the curve
presented in \cite{arplsh}, we obtain 
$\Lambda_{N_c,2N_c}^{b_1}\sim -64q$, $A_3=-\frac12$ and $B_3=-\frac12$.
For the curve  \cite{mine},
we get   
$\Lambda_{N_c,2N_c}^{b_1}\sim -108q$, $A_3=-\frac12$ and $B_3=0$.

In order to obtain the relation between the dynamical scales and the scale
parameters in the curves, let us
consider the substitution $e_i=e_i'-b\, (i=1,\cdots,N_c-1)$,
$e_{N_c}=(N_c-1)b$ and take the $b\rightarrow\infty$ limit in the 
curves (\ref{curveone}), (\ref{curvetwo}).
This corresponds to the Higgs breaking we considered in the check of 
the microscopic instanton calculation, and in fact the curve 
with $(N_c,N_f)$ reduces to that with 
$(N_c-1,0)$ after the rescaling and shift of $x$ and $y$  and the substitution
\be
\Lambda_{N_c-1,0}^{2N_c-2}=N_c^{-2}b^{N_f-2}\Lambda_{N_c,N_f}^{2N_c-N_f}.
\label{matexa}
\ee
Comparing (\ref{matexa}) with the physical matching condition (\ref{matphy})
and using the known relation 
$\Lambda_{d,2,0}=\Lambda_{2,0}$ \cite{FINPOU,itsa}, we obtain 
the relation between the dynamical scales and the scale parameters of 
the curves as\footnote{For $N_f=2N_c$ case, one notices that the square
inverses of the gauge coupling constant $1/g^2$ in the microscopic theory
and $1/g_{\mbox{ex}}^2$ appearing in the exact solution are different by a 
constant shift 
in general. This is not an inconsistency even for the scale invariant
case, because finite renormalizations might exist in general.}   
\be
\Lambda_{N_c,N_c}^{b_1}=i^{N_f}2^{-b_1/2+2}\Lambda_{d,N_c,N_f}^{b_1}.
\label{relation}
\ee

Now let us discuss the differences between the microscopic instanton 
calculation and the exact solution. 
Comparing (\ref{insres}) and (\ref{fonar}) with using (\ref{relation}),
the $u_1^{inst.}$ and the $u_1$ is the same except $U_1^{inst.}$
and $U_1$.
One can show easily that the structures of the poles are 
the same between $U_1^{inst.}$  and $U_1$, and hence the possible 
difference
is a  regular term. 
{}From the gauge invariance and the dimensional counting,
this regular term is restricted in the following form:
\be
U_1(N_c,N_f)=U_1^{inst.}(N_c,N_f)+C_{N_c}\delta_{N_f,2N_c-2}
+D_{N_c}\delta_{N_f,2N_c}\sum_{i=1}^{N_c}a_i^2.
\label{discrep}
\ee
In fact, $C_3,D_3\ne 0$ for all the above proposed curves.

These differences do not seem to lead to any inconsistencies because, 
for the $N_f=2N_c-2, 2N_c$ cases, 
the anomaly free symmetries 
certainly allow the above ambiguities in the construction of the 
curves \cite{haoz,arplsh,mine}.
There would also be some non-perturbative ambiguities in the definitions 
of the quantum operator ${\rm Tr}A^2$ in the microscopic 
theory\footnote{One would have $u^{inst.}=c(q)u\ (c(0)=1)$ and 
$u^{inst.}=u+c'\Lambda^2$ for massless $N_f=2N_c$ and $N_f=2N_c-2$ cases, 
respectively, which could explain the discrepancies in (\ref{discrep}).
For $N_{c}=2$ case, see refs.  \cite{HSDKM}}.
For $N_{c}=2$ case, this type of resolution 
has been discussed in \cite{HSDKM}.
We point out  that these differences cause some qualitative 
differences in the non-perturbative renormalization procedures  
under reductions of the systems. 
Consider for example the
Higgs breaking  $a_i=a_i'-b\, (i=1,\cdots,N_c-1)$,
$a_{N_c}=(N_c-1)b$  ($b\rightarrow \infty$) 
while $K$ matters are kept massless; $m_i=i\sqrt2 b\ (i=1,\cdots, K)$. 
Then the system with $(N_c,N_f)$ reduces to that with $(N_c-1,K)$.
When it is applied to the microscopic result (\ref{fonefull}), 
there appear divergences which cannot be absorbed into the matching 
condition (\ref{phymatcon}). 
But it can be easily shown that the divergences are
regular terms and (\ref{fonefull}) becomes consistent if the operator
$u$ is subtracted by divergent regular terms  
$u^{inst.}_{N_c-1,K}=u^{inst.}_{N_c,N_f}-(regular\ terms)$ under the reduction.
On the other hand, we do not need such a subtraction for the modulus $u$
under the similar reduction in the curves proposed in refs. 
\cite{arplsh,mine}.
Another example is the reduction in the mixed branch.
If one sets $a_{N_c}=0$ in the curve of \cite{arplsh} 
with $(N_c,N_f)$ for the massless
case, the curve reduces to the one with $(N_c-1,N_f-2)$. This can be
justified physically by assuming that the mixed branch $a_{N_c}=0$
with $q_a^i=\< q\> \delta_{N_f-1}^i\delta_a^{N_c}$,  
$\tilde{q}_i^a=\< q\> \delta_i^{N_f}\delta_{N_c}^a$ touches
the Coulomb branch at $\< q\>\sim 0$ \cite{arplsh}. 
On the other hand, the microscopic one-instanton result (\ref{fonefull})
is not consistent with naively setting $a_{N_c}=0$, and $u$ needs again
regular term shifts.
A further analysis would be necessary 
in order to clarify the physical meaning of this difference.

After completion of this work, we noticed the preprints 
\cite{HOK}, in which 
the explicit evaluations of the prepotentials of the exact solutions are
discussed for $N=2$ supersymmetric gauge theories.
 
\section*{Acknowledgments}

N.S. would like to thank H.~Kanno, 
H.~Kunitomo and M.~Sato for valuable discussions.
N.S. is supported by JSPS Research Fellowship for Young 
Scientists (No.~06-3758).
The work of K.I. is supported in part by 
the Grant-in-Aid for Scientific Research from
the Ministry of Education (No.~08740188 and No.~08211209).

\end{document}